\documentclass[aps,prd,amsmath,twocolumn,floatfix]{revtex4}

\usepackage{graphicx}
\usepackage{dcolumn}
\usepackage{bm}
\usepackage{verbatim}
\usepackage{amssymb}
\usepackage{epstopdf}
\begin{document}

\preprint{APS/123-QED}

\title{Density profiles and voids in modified gravity models}
\author{Matthew C. Martino \& Ravi K. Sheth \\
        Department of Physics and Astronomy, University of Pennsylvania, 
        209 S 33rd St, Philadelphia, PA 19104\\
        Center for Particle Cosmology, 209 S 33rd St, Philadelphia, PA 19104}
\email{mcmarti2@sas.upenn.edu, shethrk@physics.upenn.edu}

\label{firstpage}

\begin{abstract}
 We study the formation of voids in a modified gravity model in which gravity is generically stronger or weaker on large scales.  We show that void abundances provide complementary information to halo abundances:  if normalized to the CMB, models with weaker large-scale gravity have smaller large scale power, fewer massive halos and fewer large voids, although the scalings are not completely degenerate with $\sigma_8$.  Our results suggest that, in addition to their abundances, halo and void density profiles may also provide interesting constraints on such models:  stronger large scale gravity produces more concentrated halos, and thinner void walls.  This potentially affects the scaling relations commonly assumed to translate cluster observables to halo masses, potentially making these too, useful probes of gravity.  
\end{abstract}

\maketitle
\section{Introduction}
A wealth of observations, from WMAP, supernovae Ia, galaxy clustering on large scales, and cross-correlations between galaxies and the CMB, suggest that we live in a spatially flat Friedmann-Robertson-Walker universe currently dominated by either a cosmological constant or repulsive dark energy. The best fit for the dimensionless energy density parameters are $\Omega_m=0.28$ and $\Omega_\Lambda=1-\Omega_m$, within the concordance $\Lambda$CDM Model \cite{wmap5}.  However, because this requires the vast majority of the energy density in the universe to be in two unknown substances, dark matter and dark energy, there is considerable interest in alternative interpretations of what the data imply.  

The key equation of General Relativity, Einstein's equation, relates the curvature and the expansion rate of the Universe to its matter and energy content. The current paradigm is to modify the matter content of the universe, by including dark matter and dark energy, to account for observations. Instead, however, we might modify how the universe curves in response to matter, which would mean modifying our theory of gravity.  There are many ways that one could modify gravity \cite{B-D,milgrom,bekenstein,CarrollST,fr1,fr2,fr3,Cham,dgp,dgpSims,dgpPT,dgpSim2,lss,mannheim,diaferio,huReview}, but the one that we will focus on in this paper is purely phenomenological.  The assumption is that it is merely a modification to the potential, changing the form from that of standard gravity to:
\begin{equation}
\label{potentialeq}
\phi(r)=G m \frac{1+\alpha (1-e^{-r/r_s}) - \alpha (1-e^{-r/r_c})}{r}
\end{equation}
This potential (with $r_c\to\infty$ has been studied before \cite{ngp,shirata1,sealfon,fritz,shirata2} and in some sense, is an interesting limiting case of some $f(R)$ models \cite{fR,fRsims}.  In what follows, we will use $r_c\gg r_s$ but finite, following \cite{mss09}.  The precise choice of $r_c$ does not matter for any of our conclusions. 

In \cite{mss09} we studied the formation of nonlinear objects in this model -- the first study of virialized dark matter halo abundances in any modified gravity model.  In Section~\ref{voids}, we study the evolution of the structures that are in some ways the opposites of halos, namely voids.  Along the way, we show that the density profiles of objects in these models exhibit interesting departures from standard gravity models.  Section~\ref{stats} shows how we use insights from the evolution model to estimate void abundances.  Section~\ref{sims} shows that our model captures the essence of how void abundances depend on $\alpha$ and $r_s$ in numerical simulations of structure formation with this modified potential.  A final section summarizes our results.  An Appendix shows the corresponding trends for halo profiles; these potentially allow X-ray observations to provide powerful probes of modifications to gravity.  

\section{Evolution of underdensities}\label{voids}
This section describes the evolution of initially underdense spherically symmetric regions for the modified gravity model described in the previous section.

\subsection{The spherical tophat model}\label{tophat}
Following \cite{gg72}, we consider the evolution of a spherical patch of density different from the background in an expanding universe. The spherical collapse calculation begins with the statement that the force driving the acceleration is related to the gravitational potential by
\begin{equation}
 \label{F=gradPhi}
 \frac{d^2 r}{d t^2} = F = -\nabla \Phi(r).
\end{equation}
This can be integrated once to get
\begin{equation}
 \label{solveeq}
 \frac{1}{2} \left( \frac{d r}{d t} \right )^2 +\Phi(r) = C,
\end{equation} 
where $\Phi(r)$ is the integral of the potential over the mass distribution, and $C$ is the total energy of the patch, which is constant.  In standard gravity, the potential of a shell of mass $M$ is the same as that of a point mass at the center of the sphere, so $\Phi(r)$ reduces to $GM(<r)/r$.  The constant $C$ can be related to the initial overdensity and/or expansion rate of the patch:  the initial expansion rate is given by the Hubble expansion rate of the background in which the patch is embedded, namely in comoving coordinates $\dot x_i=0$, so $\dot r_i=\dot a_i x_i=\dot a_i/a_i (a_i x_i)$, so $(dr/dt)_i=H_ir_i$.  Including a cosmological constant presents no conceptual difference.  

In standard gravity one can directly solve this equation.  The solution is a cycloid, which is usually written in parametric form involving sines and cosines (but see \cite{ls08} for simple but accurate approximation).  Only objects which are sufficiently overdense initially will collapse (meaning the size of the patch has become vanishingly small) at the present time.  The critical value of the initial overdensity required for collapse today, $\delta_c$, does not depend on the initial size of the patch \cite{gg72,lc93}.  This scale independence of $\delta_c$ is a result of Birkhoff's Theorem:  the evolution of a tophat sphere is the same as that given by Friedmann's equations, so the actual size of the patch drops out.  (Indeed, for triaxial perturbations, $\delta_c$ depends on the size and shape of the initial patch \cite{smt01}.)

When gravity is modified, things are no longer so simple.  For example, when the potential is given by equation~(\ref{potentialeq}), Birkhoff's theorem no longer applies:  A particle offcenter in a uniform spherical shell will feel a force from the shell because the force no longer varies as $1/r^2$.  This has two consequences.  First, equation~(\ref{F=gradPhi}) can still be integrated once to get equation~(\ref{solveeq}), only now $\Phi(r)$ has contributions from both the internal and external mass distributions.  We can get $\Phi$ by integrating equation~(\ref{potentialeq}) of the mass distribution, about which more shortly, leaving $C$ and the initial value for $dr/dt$ to be determined.  As before, $C$ is the total energy (constant in time), and we set $(dr/dt)_i=H_ir_i$.  

Second, whereas evenly spaced concentric shells remain evenly spaced in the standard tophat model, this is no longer the case when the potential is modified.  This is most easily seen by writing $\nabla\Phi$ as the sum of three terms: that due to the Newtonian $1/r$ part, and those due to the Yukawa term from shells internal and external to the point (or shell) of interest (see Appendix~A for details).  If $\delta$ denotes the overdensity of the source shell, then the internal and external contributions are both proportional to $\alpha\delta$, but they have opposite signs.  As a result, the initial tophat perturbation develops a nontrivial density profile (see Appendix~\ref{collapse} for examples), meaning interior shells may cross one another as the initial `boundary' around them collapses.  When $\alpha > 0$, initially overdense perturbations become more centrally concentrated than when $\alpha=0$, as the contribution from the modified part of the potential pulls mass away from the boundary in both directions (provided the boundary is comparable to $r_s$).  When $\alpha <0$, then the perturbation develops a ridge at its boundary, as mass is pulled towards the boundary from both directions.  

\begin{figure}[tp]
   \centering
   \includegraphics[width=0.8\hsize]{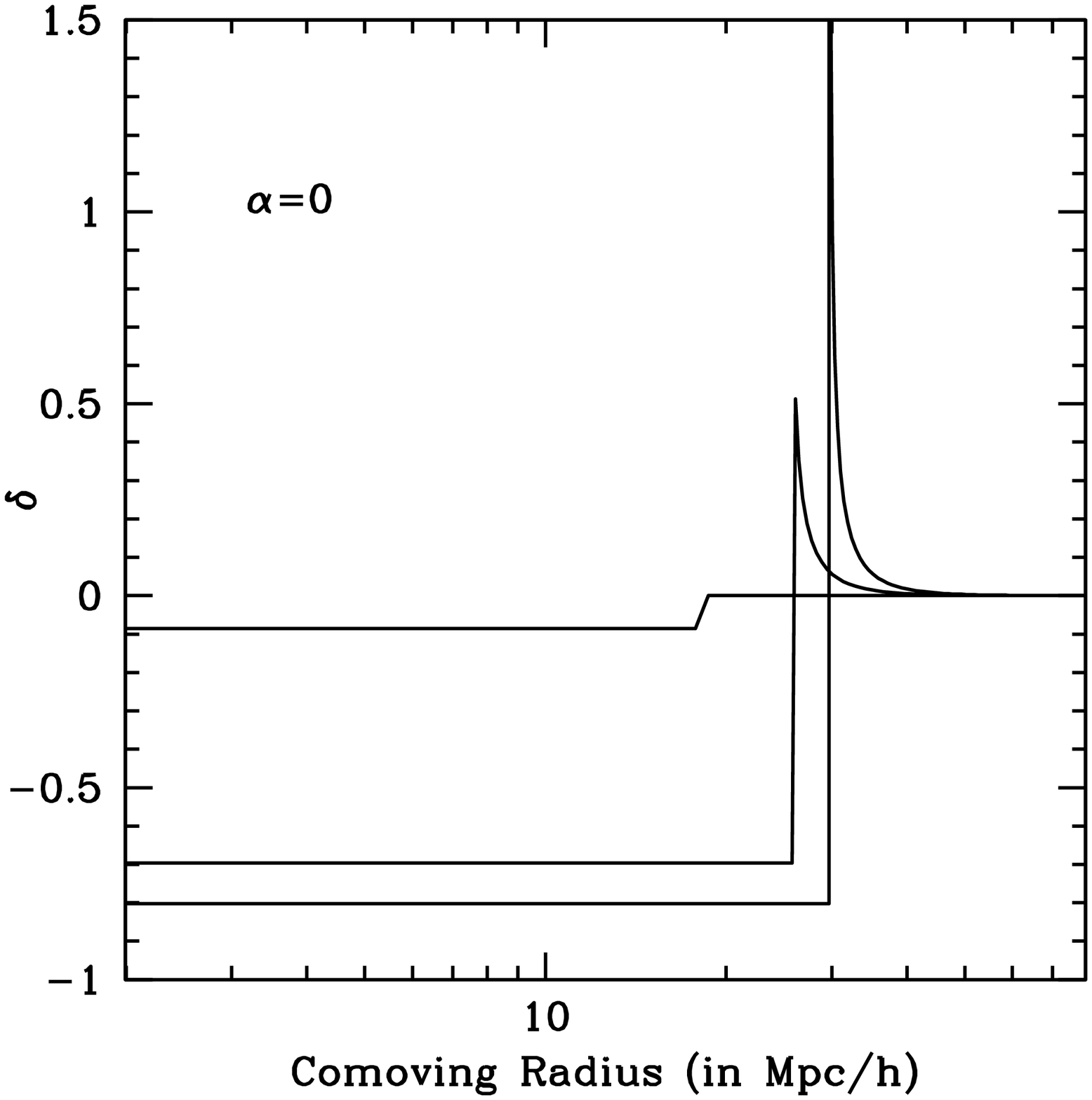}\\
   \includegraphics[width=0.8\hsize]{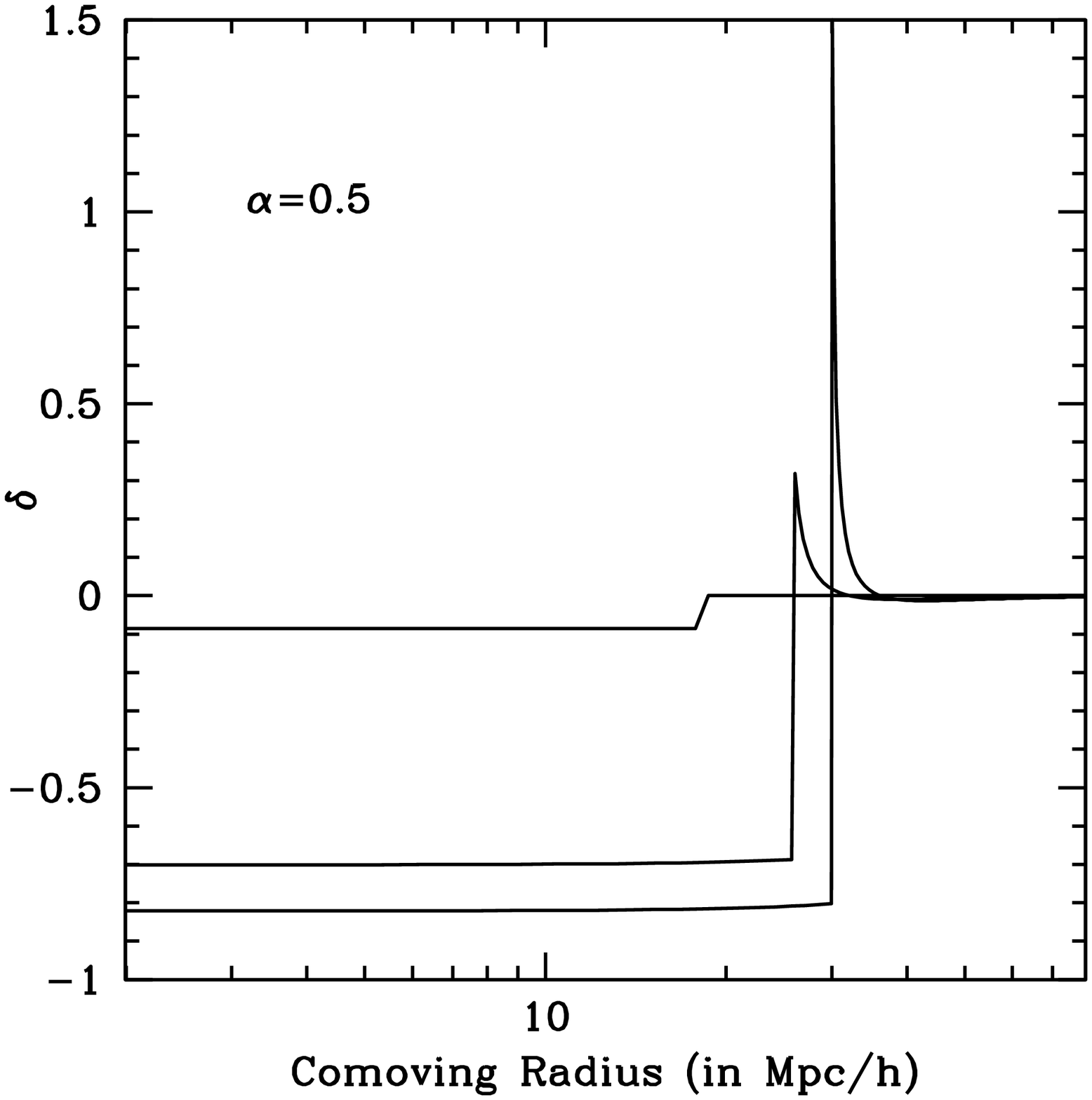}
   \includegraphics[width=0.8\hsize]{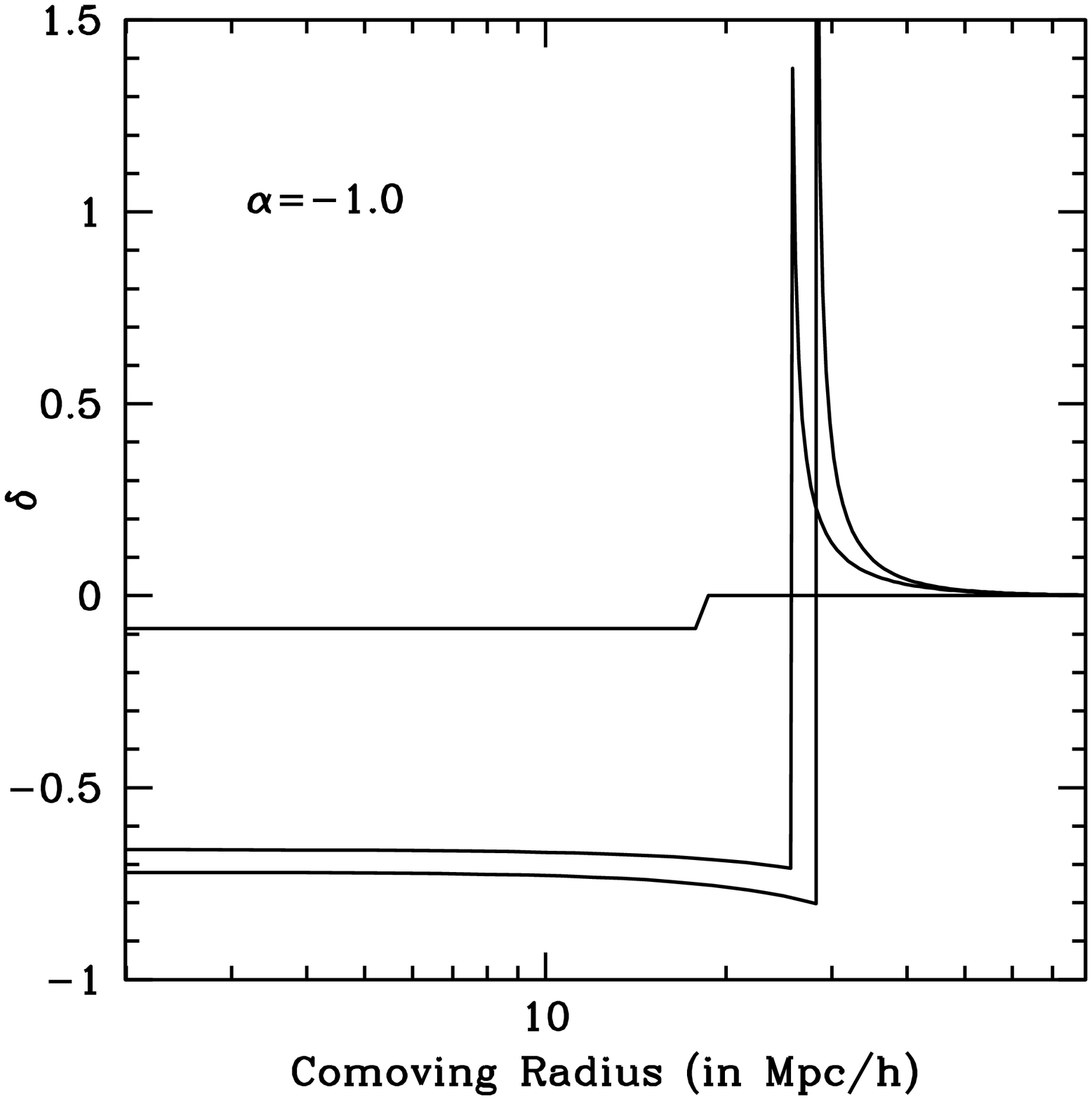}
   \caption{Evolution of the density profile of a void. The three lines are the density at the initial time, the halfway time (what would correspond to the turnaround time in halo formation), and the final time. Of note is that ridge formation occurs in all three cases, though for negative $\alpha$ the density is noticeably enhanced outside even the ridge.}
   \label{fig:density}
\end{figure}

\subsection{Density profiles}
For voids, much of the preceding discussion remains valid. In standard gravity ($\alpha = 0$), the solution is not a cycloid, but similar:  the sines and cosines are replaced by their hyperbolic counterparts (however, the simple approximation of \cite{ls08} remains accurate even in this regime).  When $\alpha\ne 0$, then the previous discussion continues to apply.  However, because the sign of $\delta$ is reversed, the dependence on $\alpha$ is also reversed.   As a result, mass moves towards the boundary of the profile from both sides when $\alpha > 0$, but away from it when $\alpha<0$.  

Figure~\ref{fig:density} shows the evolution of a number of concentric shells within which the initial density was uniform, but slightly below that of the background.  We used the methods of \cite{mss09} to calculate the evolution of the shells.  (Note, however, that whereas few shells are needed to accurately estimate the evolution of the boundary of the initial tophat perturbation, many more are needed to study the evolution of the density profile.)  When $\alpha=0$, then the expansion of the inner shells pushing against the exterior shells leads to the formation of a ridge \cite[][and references therein]{svdw04}.  This remains true when $\alpha\ne 0$, although, when $\alpha >0$ then the ridge is smaller in physical size, whereas the opposite is true for negative $\alpha$. In particular, for the $\alpha < 0$ case, the ridge extends further and trails off much more slowly.  In addition, when $\alpha\ne 0$, then the density in the interior region does not remain constant -- this is clearly more evident in the negative $\alpha$ case, but also evident in the positive $\alpha$ case.  Whereas the depedence of the interior density profile on $\alpha$ may be difficult to detect in observations, the effect on the ridge may be observable.  Of course, to really exploit this effect, one must study more realistic initial density profiles -- this is beyond the scope of our work.

\begin{figure}[tp] 
   \centering
   \includegraphics[width=0.90\hsize]{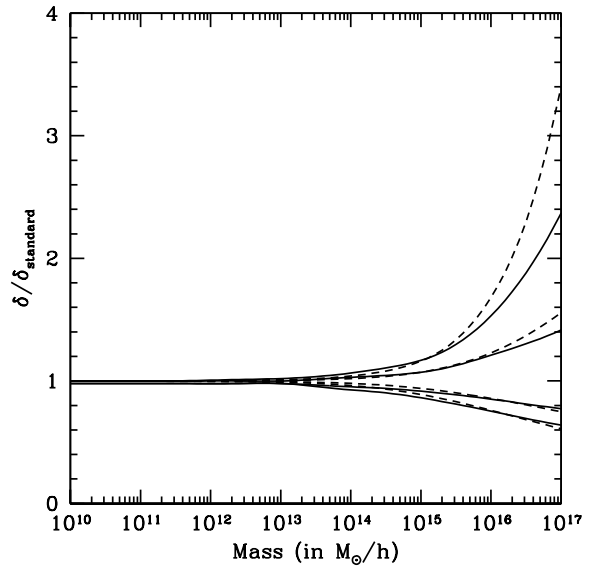}
   \caption{Ratio of initial density required for void (solid) and halo (dashed) formation at the present time to that in the standard gravity model, when the background cosmology is $\Lambda$CDM, $r_s=5h^{-1}$ Mpc and $r_c=70h^{-1}$ Mpc.  From top to bottom, curves show models in which $\alpha=-1, -0.5, 0.5$ and $1$ (note that $\alpha=0$ is standard gravity). }
   \label{fig:barrier}
\end{figure}

\subsection{Critical underdensity for void formation}
One sense in which underdensities are more difficult to study than overdensities is that the concept of a critical time, the analogue of the collapse time, is not well-defined.  While it is clear that a halo is collapsed at that time when the radius of its outermost shell reaches zero, what is the appropriate condition for undersities?  For standard gravity, one uses the condition of ``shell-crossing'' which occurs when initially interior shells first cross initially exterior shells:  the objects for which is true are about $0.2$ times denser than the background \cite{svdw04}.  Because of the ambiguity associated with shell crossing in modified gravity models, we will use this value, an underdensity of $-0.8$, to define voids even when $\alpha\ne 0$.  This critical nonlinear underdensity corresponds to some critical initial underdensity, $\delta_{v}$.  When $\alpha=0$, $\delta_v\approx 2.86$ independent of the size of the underdensity, for the same reason that $\delta_c \approx 1.686$ is independent of mass.  However, when $\alpha\ne 0$, we know $\delta_c$ depends on halo mass \cite{mss09}, and so we expect $\delta_v$ to depend on void volume.  

To find this dependence, we study the evolution of underdense patches starting from a large grid of initial underdensities and sizes.  Once we have the scale dependence of $\delta_v(R_i)$, we can relate it to a mass scale by  $M = (4 \pi /3) R_i^3 \bar\rho_i(1+\delta_i)\approx (4 \pi /3) R_i^3$, the last relation holding because $\delta_i$ is small.  Thus, we can calculate $\delta_v(M)$.  Intuitively, we expect that if, in the course of its evolution, the size of a patch never exceeds $r_s$, then we expect the large-scale modification to gravity is inconsequential.  Hence, for sufficiently small voids, which form from smaller patches in the initial field, we expect $\delta_v$ to be the same as in standard gravity.  For large voids, the effect of the modification should be stronger. For positive $\alpha$ voids should be easier to form, so we expect $|\delta_v|$ to be smaller, whereas for negative $\alpha$ the opposite should be true.  

The solid curves in Figure~\ref{fig:barrier} show that $\delta_{\rm v}(M)$ has the expected dependence on $M$ and $\alpha$.  Note that we have chosen to express the scale dependence of $\delta_v$ in terms of $M$, for ease of comparison with the scale dependence of $\delta_c$ shown in Fig.~2 of \cite{mss09}, and reproduced here as the dashed curves.  Of course, we could have expressed it in terms of the initial size $R_i$, or in terms of the final size $r_v$.  This is because we have defined voids as being 0.2 times the density of the background, making $r_v = 5^{1/3}\,R_i\approx 1.7\,R_i$:  the comoving radius of a void is 1.7 times larger than it was initially.  A close inspection of the figure will show that $\delta_v$ departs from its $\alpha=0$ value at slightly smaller mass scales than does $\delta_c$.  This is because voids expand, so smaller mass scales can eventually exceed $r_s$ in size, and so notice that $\alpha\ne 0$.

\section{Statistics of void abundances}\label{stats}
When combined with the spherical evolution model, excursion set methods \cite{bcek91,lc93,rks98,st02,ls09} are commonly used to study the abundance of virialized objects.  They can easily be extended to the study of voids \cite{svdw04}.  In short, this is done by relating the properties of such objects with the {\em initial} (rather than linearly evolved) density fluctuation field \cite{mss09}.  

\subsection{The excursion set method}
\begin{figure}[tp] 
   \centering
   \includegraphics[width=0.90\hsize]{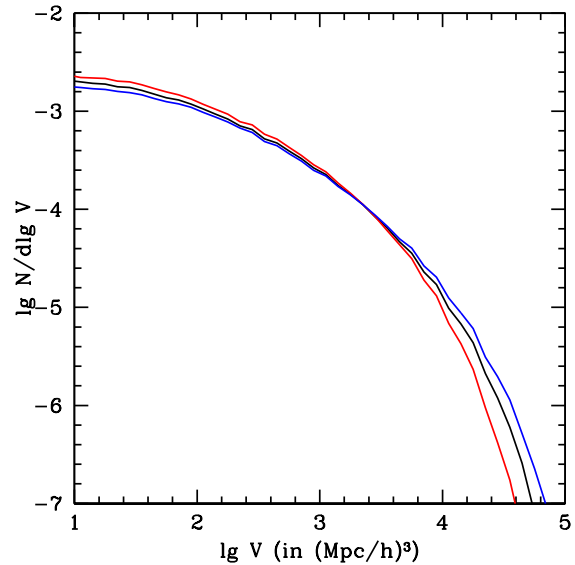}
   \caption{Mass function of voids. Black is for standard gravity ($\alpha=0$), red is for $\alpha=-1$, blue for $\alpha=0.5$. In this case, all three mass functions are calculated assuming that the initial power spectrum is the same, meaning that the power spectrum at redshift zero is different.}
   \label{fig:comparison}
\end{figure}

Now, as shown, when the potential is modified, then $\delta_v$ is no longer scale-independent.  Because it depends on mass, the relevant excursion set problem is one with a `moving' rather than `constant' barrier, so it is of the type first studied by \cite{st02}.  Figure~\ref{fig:comparison} shows the result of using this formalism to estimate the abundance of voids.  Briefly, making this estimate requires that one generate an ensemble of random walks in the $(\delta_i,S_i)$ plane, where $S_i\equiv \sigma^2_i(M)$ is the variance in the initial fluctuation field when smoothed on scale $r_i$.  Since $\sigma_i(M)$ is a monotonic function of $M$, the variables $S_i$, $M$ and $r_i$ are essentially equivalent to one another.  In particular, 
\begin{equation}
 S_i \equiv \int \frac{dk}{k}\,\frac{k^3 P_i(k)}{2\pi^2}\, W^2(kr_i),
\end{equation}  
where $W(x) = (3/x^3)\,(\sin x - x \cos x)$.  For halos, one then finds the first crossing distribution $f(S_i)dS_i$ of the `barrier' $\delta_{ci}(M) = \delta_{ci}(S_i)$.  The abundance of objects is given by 
\begin{equation}
 \frac{dn}{d\ln M}\,d\ln M = \frac{\bar\rho}{M}\,f(S_i)dS_i.
\end{equation}  

For voids all this is slightly more complicated. The relationship between $f(S_i)$ and $dn/d\ln M\,d\ln M = (\bar\rho/M)\,f(S_i)dS_i$ is unchanged, but one must account for a subtlety in the agrument.  To estimate halo abundances, one is careful to choose the {\em first} upcrossing of the collapse barrier, thus giving the largest smoothing radius which is sufficiently overdense that it will collapse today.  The larger scale environment is irrelevant: for example, if the larger scale density is negative, the end result would just be a halo which formed inside a void.  With voids, though, one must worry about underdense regions which are surrounded by large scale overdensities which would have collapsed by the present time -- the void-in-cloud problem.  Namely, if on some scale the random walk crosses below the critical density for void formation, but on a larger scale it crossed above the barrier for halo formation, then this would be a void which was crushed as the region around it collapsed.  Therefore, such walks should not contribute to the estimated void abundance.  As a result, rather than a one barrier problem, we instead have a two barrier problem \cite{svdw04}: $\delta_v$ must be crossed {\em before} $\delta_c$, is crossed.  (A more conservative version of this argument sets the barrier associated with the collapsing object to equal that for turnaround rather than full collapse.)
 
 Because there are two barriers, the first crossing distribution has a peaked shape, with the distribution being cut off at both high and low masses, on the high end due to the rarity of such extreme underdensities, and at the low end, because $\delta_c < |\delta_v|$.  In large part, we will not be interested in the intricacies of which halo formation barrier we should use, as this largely affects only the low mass end of the mass function, where the modification to gravity is smaller.

Figure~\ref{fig:comparison} shows the resulting void mass functions for standard and modified gravity.  (We do not show the extremely low mass regime, which is most sensitive to the void-in-cloud cutoff.)  Notice that when $\alpha$ is positive there are more large voids and fewer small voids, whereas the opposite is true when $\alpha$ is negative.  This is from two effects: one is that any given walk tends to cross later when gravity is weaker ($\alpha<0$) as the barrier is lower (recall $\delta_v$ is negative) and so voids tend to be shifted towards lower mass. The second effect is that because it is harder to form high mass halos in weaker gravity, some walks that cross $\delta_c$ in stronger gravity may not cross $\delta_c$ for weaker gravity, however due to the fact that they are near $\delta_c$ at high mass, it is still difficult for such a walk to cross $\delta_v$ even at low mass. The first effect is likely more significant, but both contribute to the difference at low mass.

\subsection{Initial conditions}

\begin{figure}[tp]
  \centering
  \includegraphics[width=0.8\hsize]{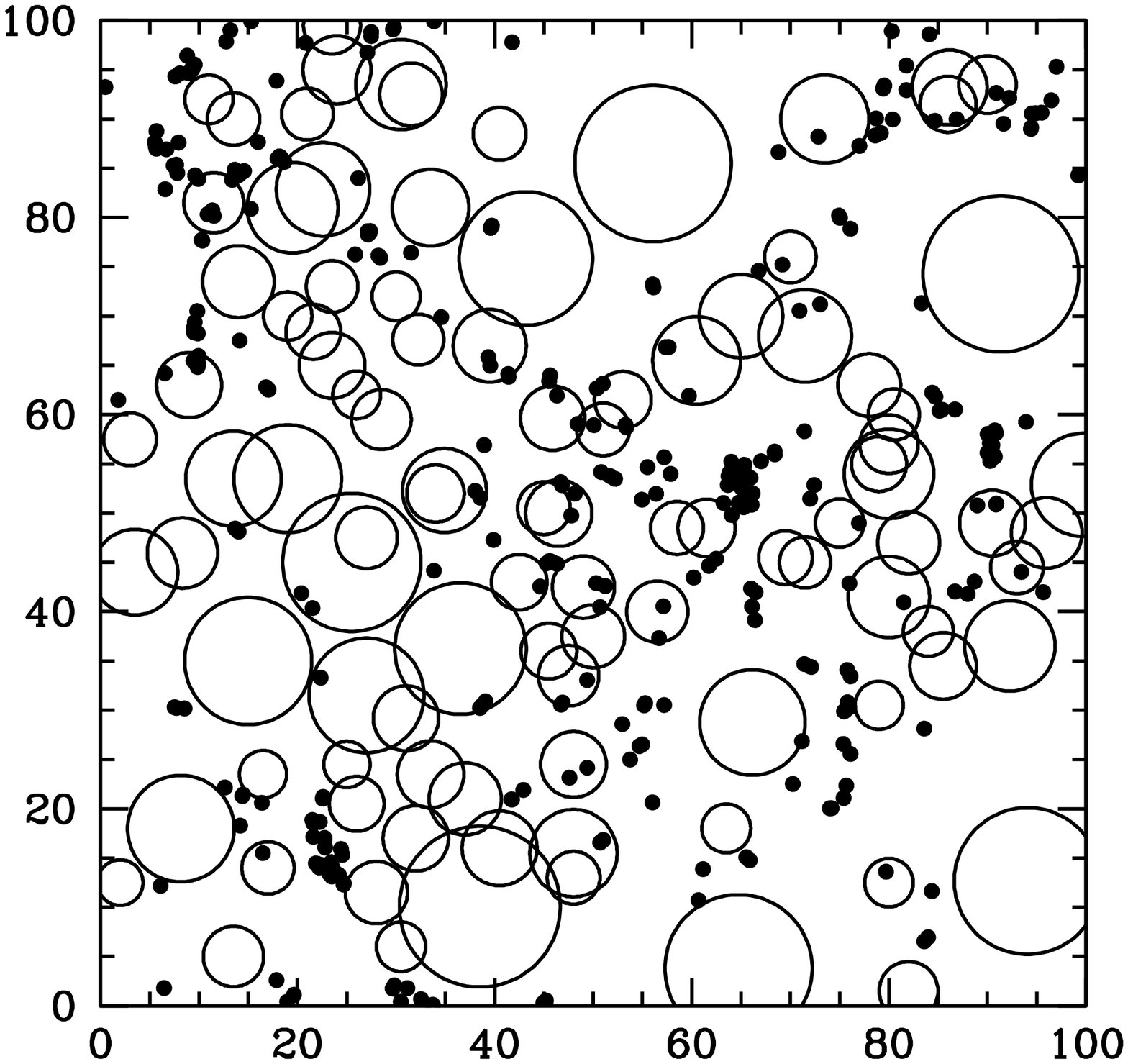}
  \includegraphics[width=0.8\hsize]{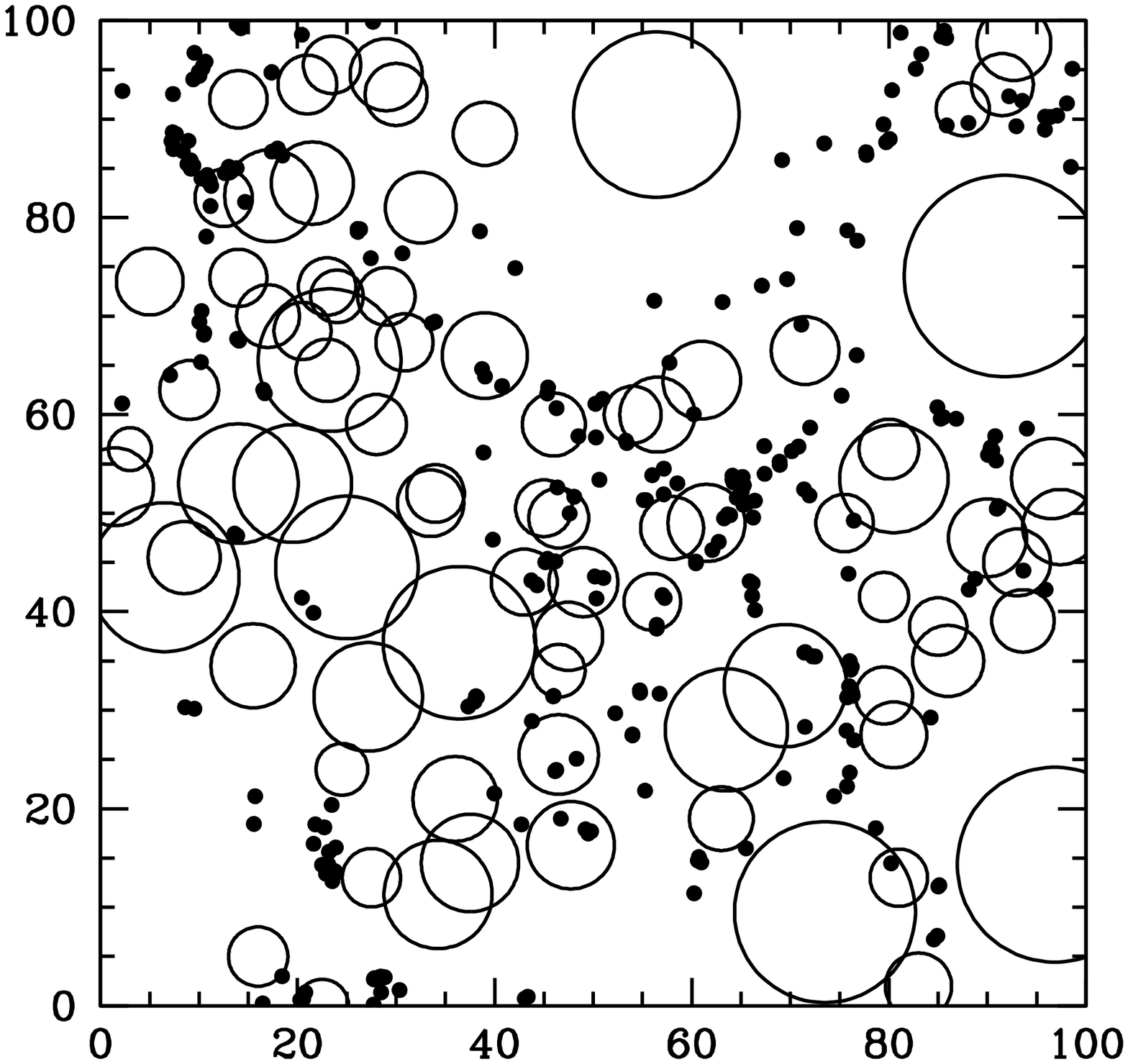}
  \includegraphics[width=0.8\hsize]{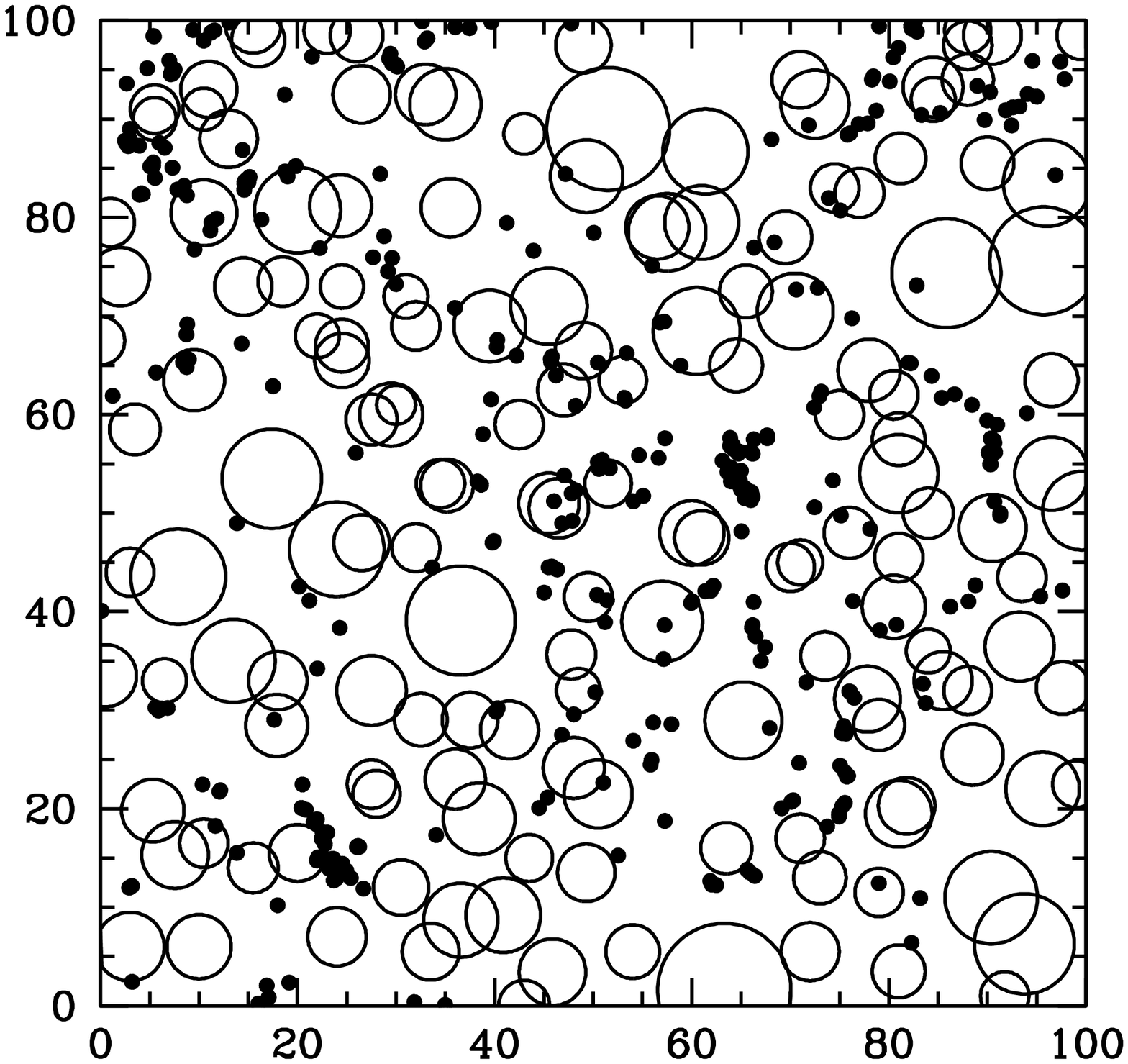}
  \caption{Spatial distribution of halos and voids in three realizations of the same initial conditions evolved to the present time using different values of $\alpha$.  Each panel shows voids and halos whose centers lie in a slice that is $6h^{-1}$Mpc thick:  top panel shows halos and voids in a standard gravity run ($\alpha = 0$), middle is for $\alpha = 0.5$, and bottom for $\alpha=-1.0$.  The halos become more strongly clustered and the voids larger as $\alpha$ increases.}
  \label{fig:pic}
\end{figure}

One thing to note about the histograms in figure~\ref{fig:comparison} is that the difference between the standard gravity and modified gravity void mass functions comes from two sources. Perhaps the more obvious is that modifying gravity modifies the behavior of individual structures, changing how quickly they form. This information is encapsulated in the barrier shape, and hence is accounted for using excursion set methods.  The result of a shifted barrier is a shift in the function $f(S)$, the first crossing distribution. The other source is the modification to the power spectrum.  In the study above, we assumed that the initial power spectra were the same for all the models -- what is usually called CMB-normalization.  However, because gravity is modified, the late time power spectra (in both linear and nonlinear theory) depend on $\alpha$ (and $r_s$).  So now there are two other possible cases. We could consider modified gravity but with the initial conditions chosen so that the $z=0$ power spectrum is that of standard gravity, or we could consider standard gravity but with the initial conditions modified so that the $z=0$ power spectrum is that of modified gravity.  In excursion set terms, either of these choices affects the relationship between $M$ and $S$.

For halos, whether one chooses to match the initial power spectrum or the final power spectrum matters \cite{mss09}, because it changes the mapping between $S$ and $M$ (or $R$), and this mapping affects two steps of the calculation:  the transformation of $\delta(M)$ into $\delta(S)$, and the relation between $f(S)$ and $n(M) dM$.  The same is true for voids so, in what follows, we will consider some of these other possiblities as well.  

\section{Comparison with simulations}\label{sims}
We now compare our spherical evolution predictions for void abundances with measurements in the simulations of \cite{fritz}.  These simulations followed the evolution of $128^3$ particles in a periodic box of size $100 h^{-1}$Mpc, for various choices of $\alpha$ and $r_s$.  In all cases, the background cosmology was flat $\Lambda$CDM with $\Omega=0.3$, and the particle mass was $1.1\,\times\,10^{10} M_\odot$.  The $\alpha=0$ simulation, with standard initial conditions has $\sigma_8=1.0$ at $z=0$.  The corresponding runs for $\alpha=0.5$ and $\alpha=-1$ have $\sigma_8=1.10$ and $0.84$ respectively.  Following our discussion of how the halo and void counts depend on the shape and normalization of the initial power spectrum, we also study results from $\alpha=0$ simulations in which the initial power spectrum was modified so that, at $z=0$, it has the same shape as the two $\alpha\ne 0$ cases ($\sigma_8=1.10$ and 0.84).

\begin{figure}
   \centering
   \includegraphics[width=0.90\hsize]{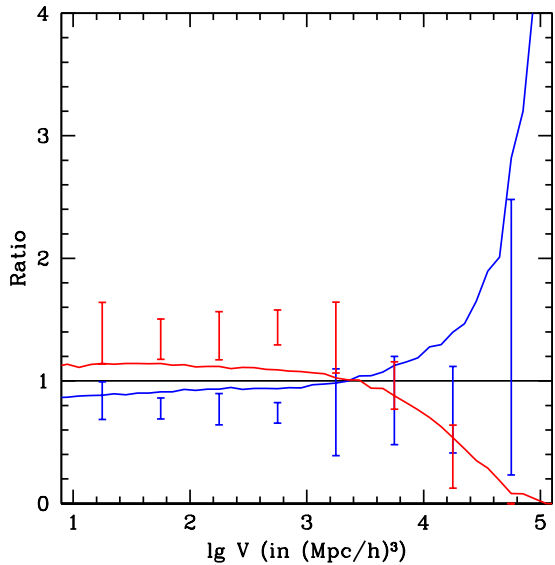}
   \caption{Ratio of mass function of voids to that in standard gravity, when the initial power spectrum is the same.  Red is for $\alpha= -1$, blue for $\alpha = 0.5$. The histograms (TO BE ADDED) are the results from the simulations, the solid lines are the results from theory.}
   \label{fig:ratio}
\end{figure}

\begin{figure} 
   \centering
   \includegraphics[width=0.90\hsize]{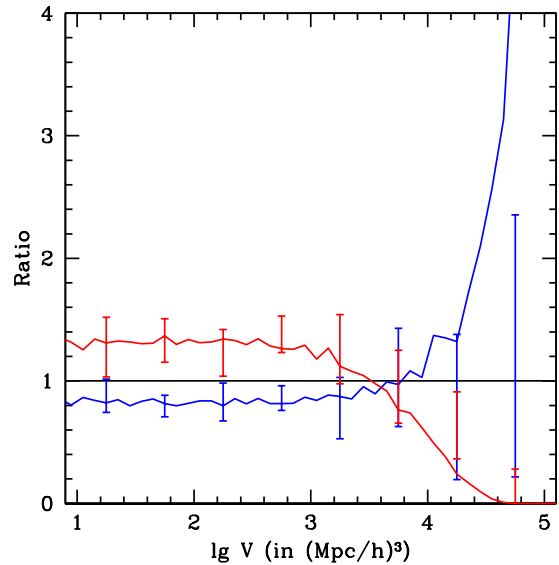} \\
   \includegraphics[width=0.90\hsize]{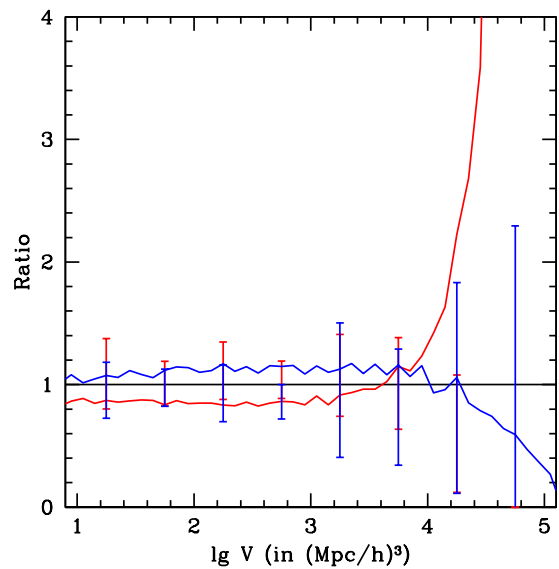}
   \caption{Ratio of mass function of voids in standard gravity with modified ICs to standard gravity with standard ICs.  Top:  The ICs are chosen so that at $z=0$, the simulations had the same power spectrum as the modified gravity simulations.  Bottom:  ICs for the $\alpha=0$ runs have been tuned to produce the same power spectrum at $z=0$ (this tuning is different for the two values of $\alpha$ shown).  In both panels, red is for $\alpha = -1$, blue for $\alpha = 0.5$. }
   \label{fig:stanmodtostan}
\end{figure}


\begin{figure} 
   \centering
   \includegraphics[width=0.90\hsize]{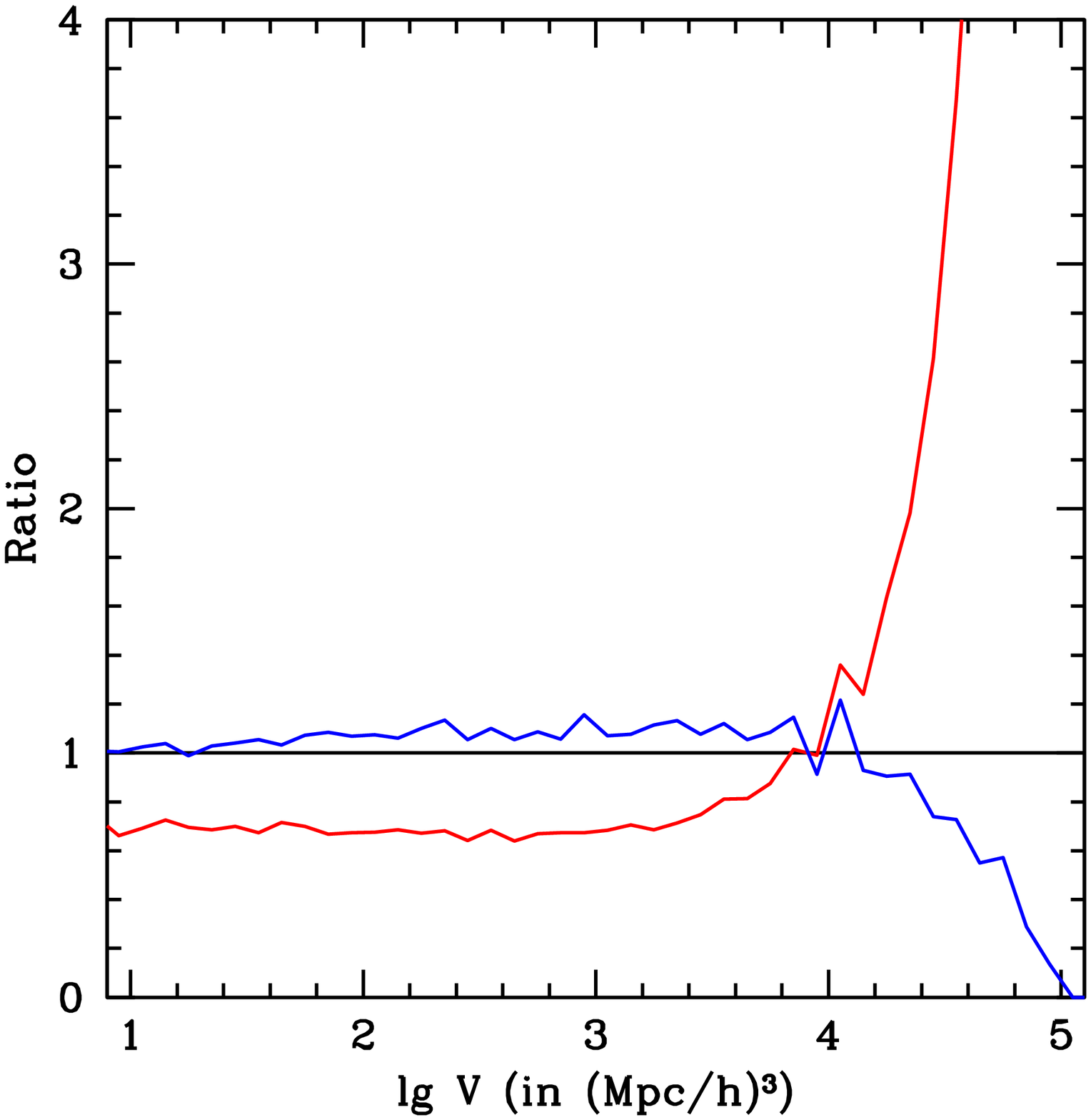}
   \caption{Ratio of void counts in modified gravity simulations to those in standard gravity.   Here, the standard gravity runs use standard initial conditions, but the modified gravity runs use different initial conditions, chosen so that the $z=0$ power spectrum is the same as it would be for standard gravity.  Red is $\alpha = -1$, blue is $\alpha = 0.5$.  }
   \label{fig:ratiomodICmodgravtostan}
\end{figure}

The simulations were analyzed using the void finder of \cite{joergVF}.  Figure~\ref{fig:pic} shows the halos and voids found in three runs, all with the same iitial conditions.  Notice that the voids are largest when $\alpha=0.5$ and smallest when $\alpha=-1$, in qualitative agreement with expectations.  Since the $\alpha=0.5$ run has more large scale power at $z=0$ this is not surprising.  Figure~\ref{fig:ratio} shows the results of a more quantitative comparison:  although qualitatively similar, the predicted effects (solid lines) for large voids are larger than we find in the simulations, and the effect on small voids is smaller than is observed.  

We explore the dependence on choice of initial conditions in two steps.  The top panel in Figure~\ref{fig:stanmodtostan} shows the ratio of void counts in two standard gravity runs, one with initial conditions modified to produce the same $z=0$ power spectrum as $\alpha\ne 0$, and the other with the standard $\sigma_8=1$ initial conditions (note that the modification to the initial conditions is different for the two values of $\alpha$ shown).  This plot is qualitatively similar to the previous one, because stronger large-scale gravity ($\alpha>0$) is qualitatively like having more large scale power, but the differences between these two plots shows that the result of modifying gravity is {\em not} degenerate with modifying the shape and normalization of the initial power spectrum.  The bottom panel shows this information slightly differently:  here, the void counts in the $\alpha\ne 0$ run are ratioed to the counts in the $\alpha=0$ run in which the initial conditions were tuned to produce the same $z=0$ power.  The fact that the ratios are not unity implies that there is more information in the void size distribution than in the power spectrum itself.

Finally, Figure~\ref{fig:ratiomodICmodgravtostan} shows the ratio of void counts when it is the  initial conditions in the the modified gravity runs which have been tuned to produce the same $z=0$ power spectrum (rather than tuning the $\alpha=0$ initial conditions).  Unfortunately, we do not have simulations of this case, but we again see that the ratio is predicted to differ from unity, indicating that modifications to gravity are not degenerate with changes to the initial conditions.  

\section{Conclusions}
 We studied a particular modification to large-scale gravity, which has two free parameters:  a scale $r_s$, and an amplitude $\alpha$ (equation~\ref{potentialeq}).  Previous work has shown that halo abundances can be modeled well enough \cite{mss09} to use them to constrain such models.  We present an example of this in Appendix~\ref{alphars}.  We argued that since voids are so large, and almost fill space, they too can be useful probes. In particular, the study of the formation of large voids provides a useful counterpart to the the study of the formation of massive halos because both probe the nature of gravity on large scales.  This is timely, given that large, statistically representative void samples are only just becoming available \cite[e.g.][]{voidp,voids}.  

 Whether stronger large-scale gravity produces more or fewer large voids depends on how one normalizes the models.  When normalized to have the same power spectrum at early times (the most common choice), stronger gravity produces more large voids (Figure~\ref{fig:comparison}).  This case also produces more massive halos, so the net result is qualitatively like having standard gravity with more large scale power (Figure~\ref{fig:pic}).  Although there are quantitative differences which allow one to distinguish between a larger normalization and modified gravity (Figures~\ref{fig:ratio}--\ref{fig:ratiomodICmodgravtostan}), we note that this fact alone -- a mismatch between the CMB- and later-time determinations of the amplitude of the power spectrum on 10~Mpc scales when standard gravity is assumed -- are generic signatures of modifications to large-scale gravity.  Note that this particular signature has the same sign for halos and for voids -- stronger large scale gravity means more massive voids and more large voids.  In contrast, if gravity is standard but the initial conditions were non-Gaussian, then halo and void abundances are modified in qualitatively different ways, at least for the local non-Gaussian initial conditions of current interest \cite{lsd09}.  

 However, if normalized to have the same power at $z=0$, the dependence of void (and halo) abundances on whether large-scale gravity is stronger or weaker depends on how one chooses to do this.  If this is done by increasing (decreasing) the initial power spectrum in models with $\alpha<0$ ($\alpha>0$), then one still expects the stronger gravity models to produce larger voids (Figure~\ref{fig:ratiomodICmodgravtostan}), although the predicted abundances are quantitatively different.  However, if one modifies the standard gravity initial conditions to match those of the $\alpha\ne 0$ models (increase/decrease initial large scale power to match $\alpha>0$/$\alpha<0$), then one predicts more (fewer) large voids when $\alpha<0$ ($\alpha>0$) relative to the $\alpha=0$ case.  This dependence on how the models were normalized is qualitatively similar to the trends seen for massive halos \cite{mss09}.  

 We presented a method for estimating these effects which is in good qualitative greement with the simulations, but there are quantitative differences.  However, the agreement is not as good as it was for describing halos.  Larger simulations are required to determine if this is due to the relatively small size of the simulation boxes available to us, or to some more fundamental problem with our analysis (see \cite{smt01,ls09} for discussion of the drawbacks of the excursion set approach).  

\begin{figure}[p] 
   \centering
   \includegraphics[width=0.7\hsize]{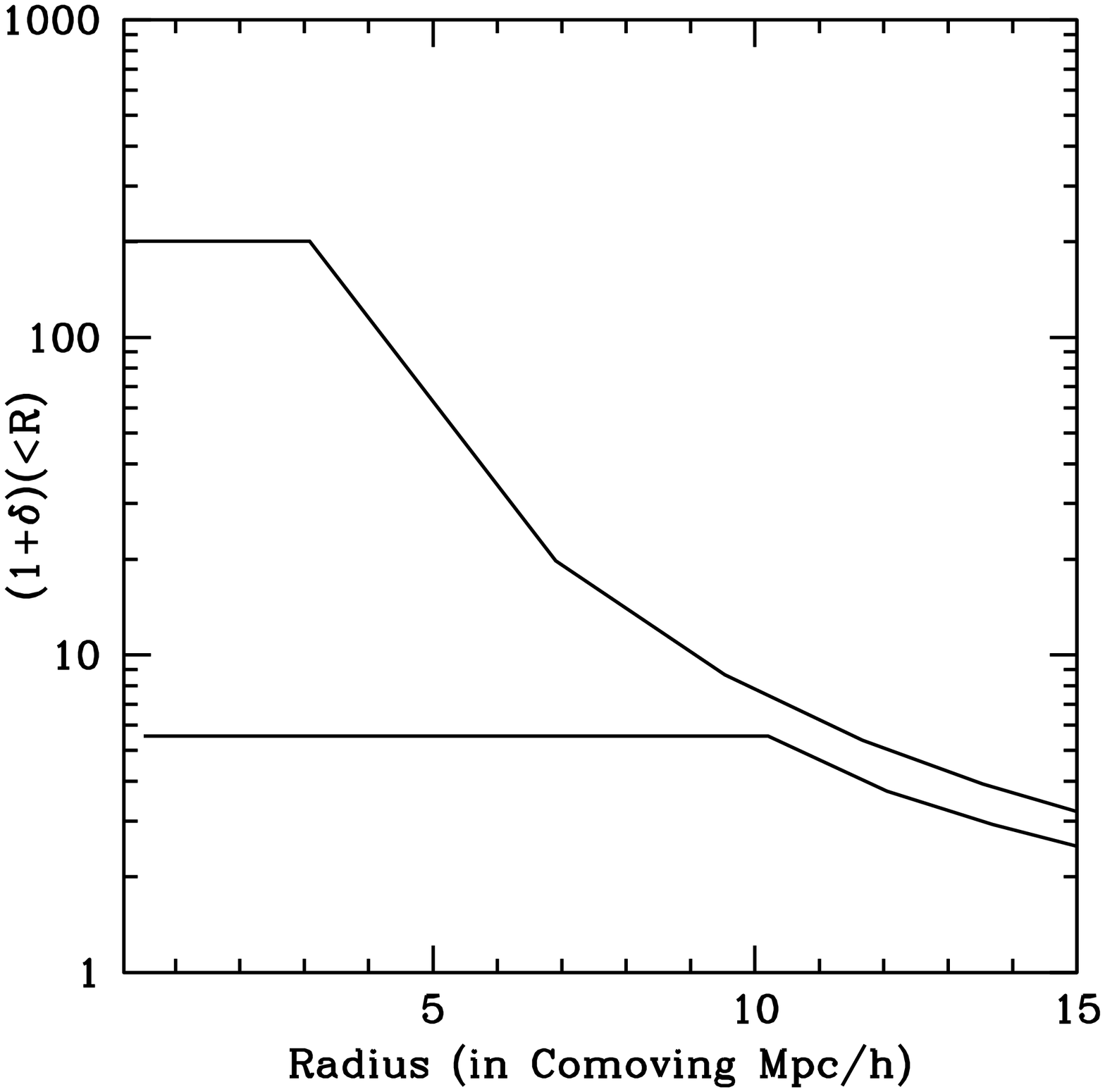} 
   \includegraphics[width=0.7\hsize]{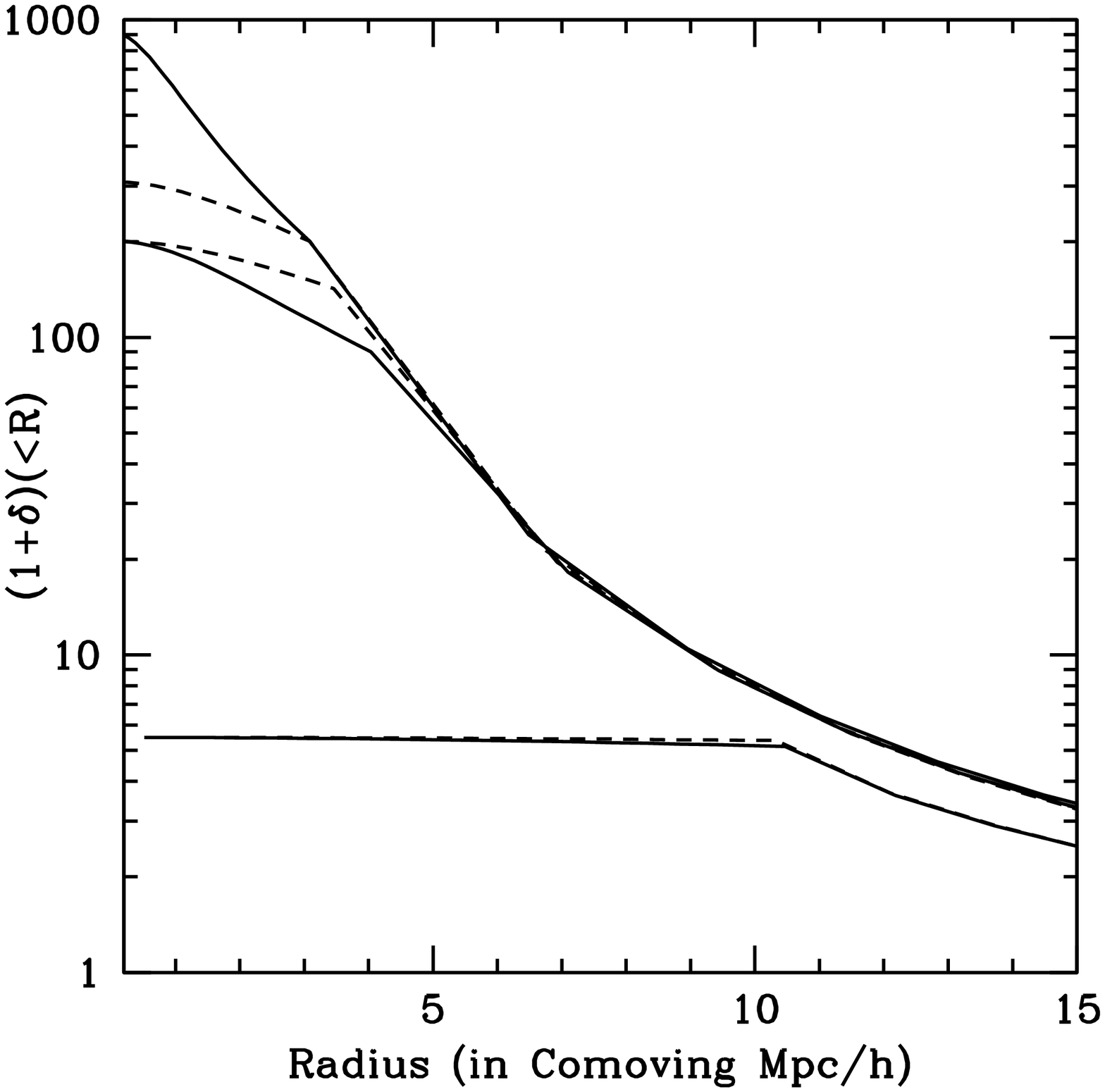} 
   \includegraphics[width=0.7\hsize]{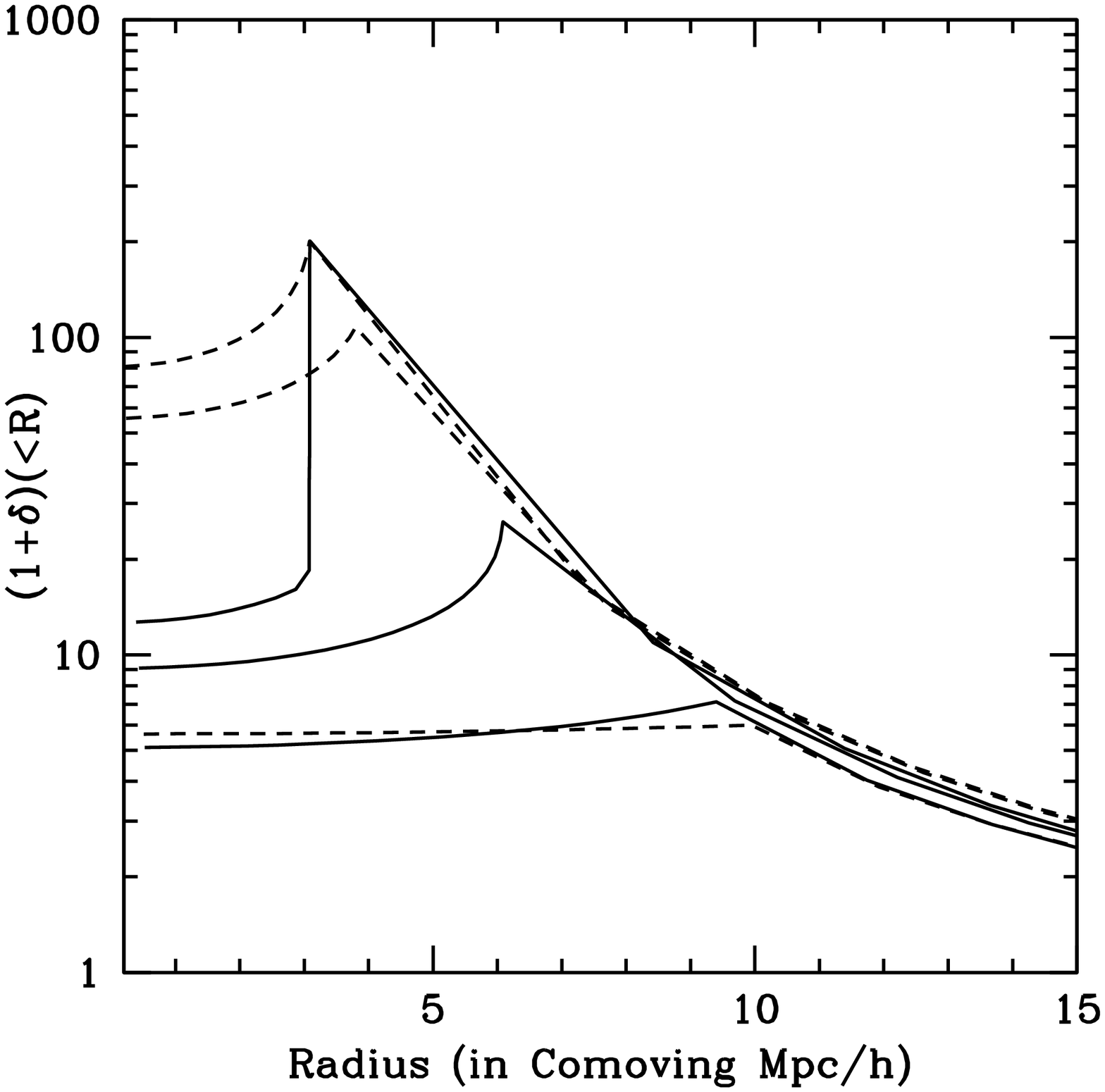}
   \caption{Evolution of the enclosed density of initially tophat overdense regions in standard (top) and modified gravity (middle and bottom panels show results for  $\alpha=0.5$ and $\alpha=-1.0$ respectively). In the modified gravity plots, solid and dashed curves are for $r_s=5h^{-1}$Mpc and $r_s=10h^{-1}$Mpc respectively. In all cases, the mass of the final halo is $10^{15.3}h^{-1} M_\odot$, and the density profile is shown (from bottom to top) at turnaround, when a shell reaches a local density of $200\,\rho_b$, and when shell within which the enclosed mass is $10^{15.3}h^{-1} M\odot$ reaches an enclosed density of $200\,\rho_b$.  (For standard gravity these last two conditions occur simultaneously.)  Increasing $r_s$ leads to halo profiles that are more similar to standard gravity.}\label{fig:densityhalos}
\end{figure}

 We also showed that the density profiles of voids and halos may also provide interesting probes of modified gravity.  When the initial profile is a tophat, then the evolved halo profile in the $\alpha<0$ case generally has a cusp at the virial radius (mass flows towards the boundary), whereas when $\alpha>0$ the halos are more centrally concentrated (Figure~\ref{fig:densityhalos}).  The structure of the void walls also depends slightly on $\alpha$ (Figure~\ref{fig:density}).  We provided tentative evidence that the profiles of halos in simulations do depend on $\alpha$ (Figure~\ref{simprofs}), and hope that these initial measurement motivate further study of this interesting problem.  It will be interesting indeed if these trends persist in simulations with better mass and force resolution, because it is conceivable that effects of this magnitude will soon be measured by weak lensing surveys.  The more dramatic effects associated with initially tophat profiles potentially provide an interesting X-ray signature of modified gravity -- it would be interesting to simulate such models using SPH codes.  

 In conclusion, we expect our results will prove useful in studies which use the large scale distribution of galaxies, and the structure of galaxy clusters, to constrain large scale modifications of gravity.  

\section*{Acknowledgments}

We would like to thank Fritz Stabenau for providing the outputs from his simulations in electronic form, and Joerg Colberg for running his void finder algorithm on these outputs.  This work was supported in part by NSF-AST 0908241.

\appendix
\section{Halo density profiles}\label{collapse}
A complementary study to the one in the main text regarding the evolution of the density profile of voids is a study of the evolution of halo density profiles. Figure~\ref{fig:densityhalos} shows one such study.  In this case, the three objects all started as tophats with the same initial overdensity, but different values of $\alpha$.  The solid and dashed curves show the density profile for two different values of $r_s$ and at three different times, one corresponding to turnaround, one to when one shell reaches a density of 200 times the background, and the final to when the density enclosed in the outer shell reaches 200 times the background.  Since gravity has a different strength in the three cases, and the patches all started with the same overdensity, the final collapse happeds at different times. For positive $\alpha$, objects collapse faster, whereas for negative objects collapse slower, with the difference decreasing as $r_s$ increases.  This is why the effect is noticeable earlier for negative $\alpha$: the turnaround time is later than for standard gravity, so the deviation is greater.

\begin{figure} 
   \centering
   \includegraphics[width=0.8\hsize]{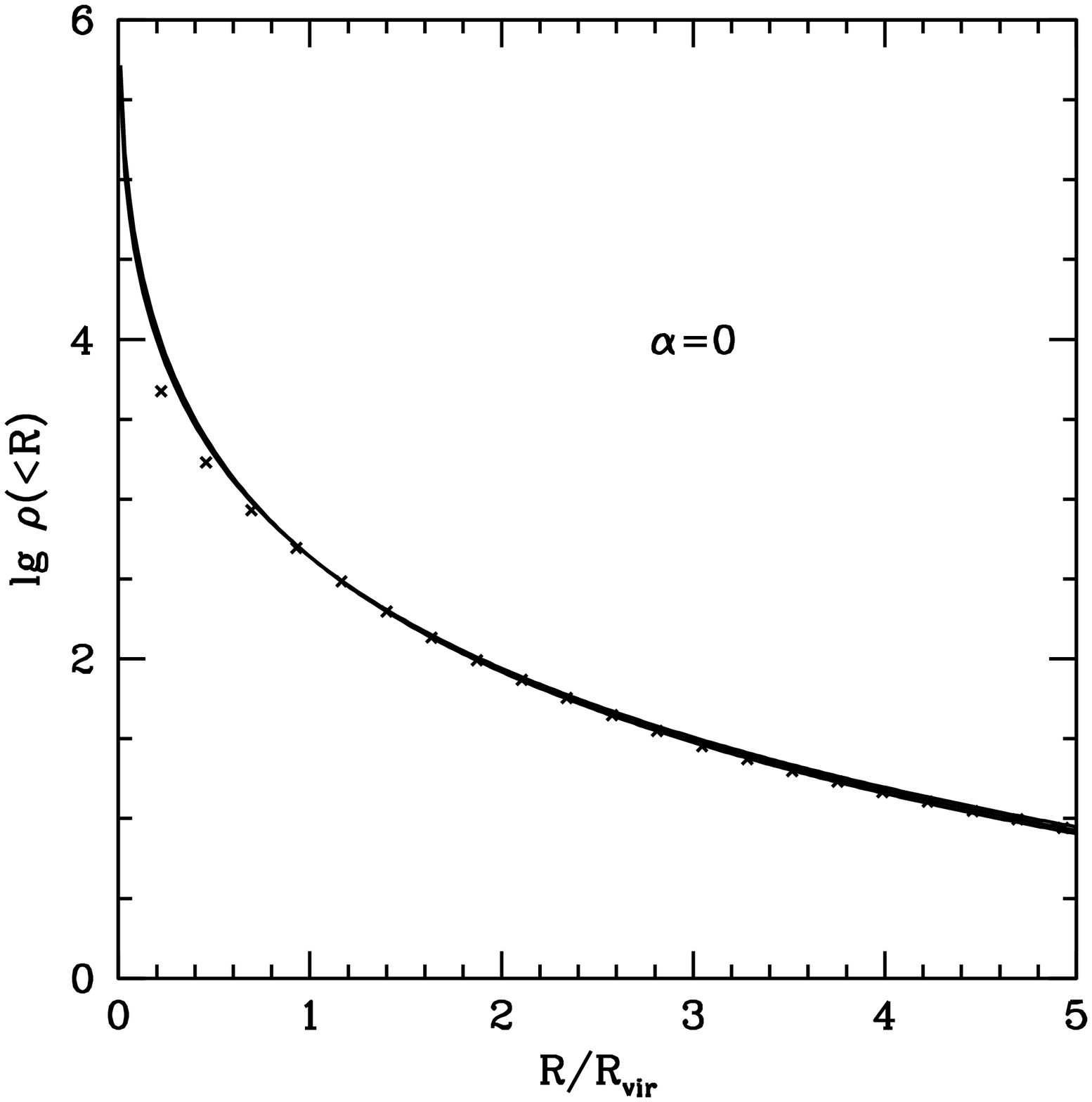} 
   \includegraphics[width=0.8\hsize]{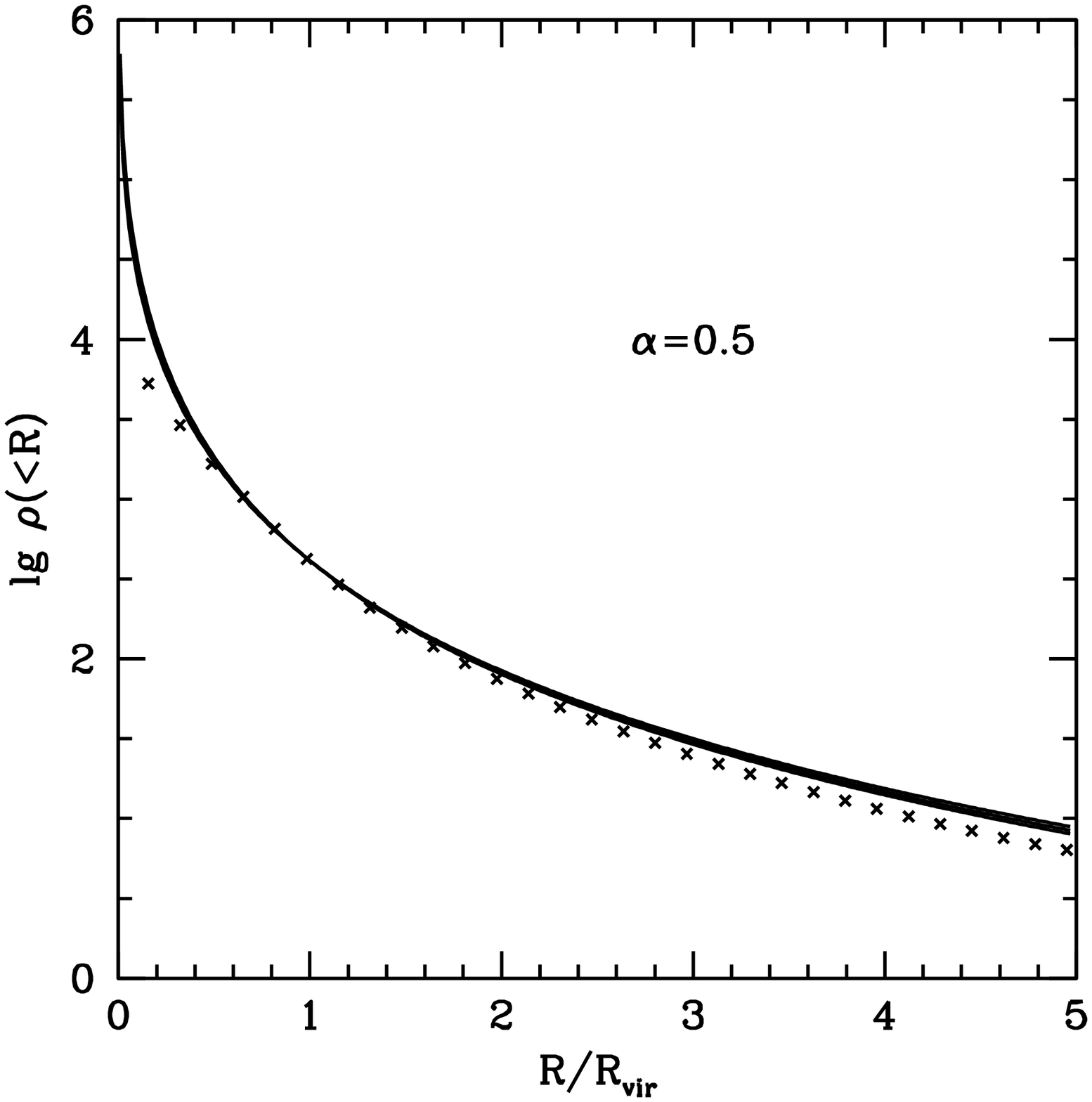}
   \includegraphics[width=0.8\hsize]{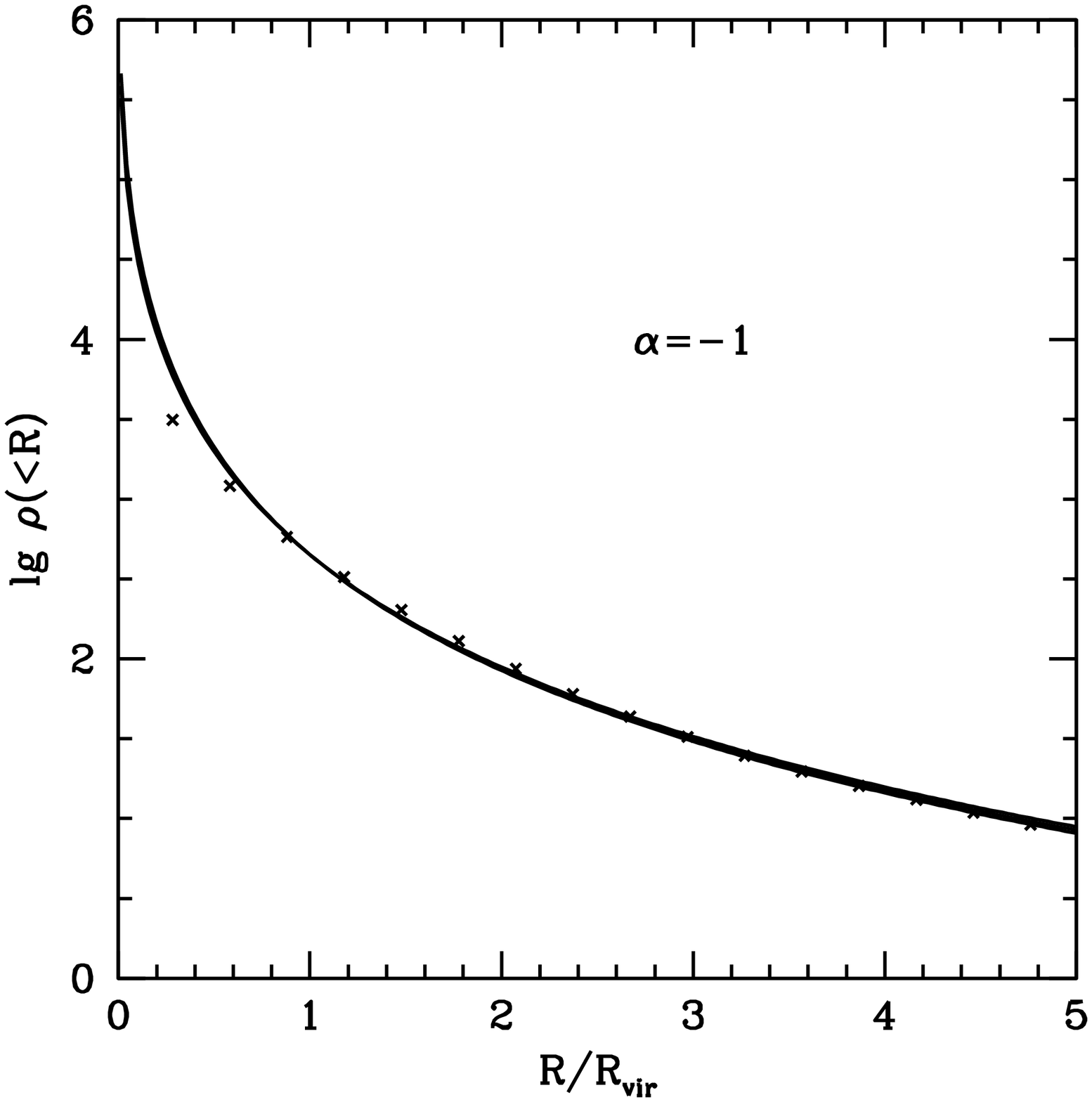} 
 \caption{Profiles of the enclosed density for the most massive halos in 
          the simulations started from identical initial conditions 
          but with $\alpha=0$ (top), $alpha=0.5$ (middle) and 
          $\alpha=-1$ (bottom).  }
 \label{simprofs}
\end{figure}

The evolved profiles in the two modified gravity cases are strikingly different from standard gravity, with the formation of sharp spikes at the edge of the halo in the negative $\alpha$ case, and the formation of a large central peak in the positive $\alpha$ case. Whether these profiles would hold through virialization is unknown, but since the effect is so strong during the period before collapse it seems likely that the profiles of virialized haloes would be affected.  If so, then halo profiles might provide new interesting constraints on $\alpha$.  However, to place such constraints, one would have to extend our analysis to more realistic initial profiles which do not have sharp edges -- although it is likely that the formation of the cusp at the boundary when $\alpha<0$ will remain (see Figure~\ref{simprofs} and related discussion). Note that for larger $r_s$, the difference from standard gravity is smaller, because the perturbation spends essentially all its time at scales which are smaller than $r_s$.

These trends are perhaps most easily understood by writing the potential due to an extended source at scales that are beyond its boundary.  The potential at $r$ due to a top perturbation which contains mass $M$ within radius $R$ can be written as the sum of the standard gravity piece plus a term which arises from the modification:  
\begin{eqnarray}
 \Phi(r) &=& \frac{GM}{r} - \alpha\, \frac{G\delta M}{r}\,
              \Bigl[{\rm e}^{-r/r_s}\, {\cal F}(R/r_s) \\
         && \qquad\qquad\qquad\qquad - {\rm e}^{-r/r_c}\, {\cal F}(R/r_c)\Bigr]
         \nonumber
\end{eqnarray}
where $\delta M \equiv M - \bar\rho 4\pi R^3/3$ and 
\begin{equation}
 {\cal F}(x) = \frac{x\,\cosh(x) - \sinh(x)}{x^3/3}.
\end{equation}
Note that ${\cal F}\ge 1$ for all $x$.  However, ${\cal F}\to 1$ when $x\ll 1$, so it is easy to see that when $r_s\gg R$ then $\Phi$ looks just like it does standard gravity.  If the overdensity profile is a power-law of slope $-1$ within $R$, so 
 $\delta M = 2 \pi R^3$, 
then the expression for $\Phi$ remains true, but
 ${\cal F}(x)\to [\cosh(x)-1]/(x^2/2)$:  
again, ${\cal F}\ge 1$ for all $x$, and ${\cal F}\to 1$ when $x\ll 1$.  
In practice, we are most interested in the regime in which $R$ is smaller than a few times $2r_s$, for which the difference between these form factors is small.  This means that one can approximate the force outside the perturbation by assuming it is a tophat, and not worrying about its internal structure.  In fact, this suggests that one can solve for the evolution of the boundary of the perturbation without worrying about the fact that the internal structure is changing; in essence this is why the analysis in \cite{mss09} was so straightforward.  



The evolution of the profile within the perturbation is more easily studied using the force, which is $\nabla_r \Phi(r)$, and can be split into a Newtonian part, plus terms due to the modification.  These in turn can be split into the contributions from shells internal and exterior to $r$.  If we write the mass overdensity in the $i$th shell as
 $\delta M_i = \bar\rho(a)\, \delta_i\, 4\pi r_i^2\, dr_i$ then 
\begin{equation}
 \nabla_r\Phi = \alpha\,\frac{G\,\delta M_i}{r_i^2} \frac{r_i}{r}\, 
                \frac{F_i(r_s) - F_i(r_c)}{2}
\end{equation}
where 
\begin{eqnarray}
 F_i(r_x) &=& \left(1 + \frac{r_x}{r}\right)\,
       \left({\rm e}^{-|r-r_i|/r_x} - {\rm e}^{-(r+r_i)/r_x}\right)\nonumber\\
 &=& 2\,\left(1 + \frac{r_x}{r}\right)\, {\rm e}^{-r/r_x}\,\sinh(r_i/r_x) 
\end{eqnarray}
when $r_x\le r$ (i.e., from the internal shells) and 
\begin{eqnarray}
 F_i(r_x) &=& \left(\frac{r_x}{r} - 1\right)\,{\rm e}^{-|r_i-r|/r_x} 
            - \left(1 + \frac{r_x}{r}\right)\,{\rm e}^{-(r+r_i)/r_x}\nonumber\\
          &=& -2\,\frac{r/r_x \cosh(r/r_x) - \sinh(r/r_x)}{r/r_x}\,
              {\rm e}^{-r_i/r_x} \nonumber\\
          &=& -\frac{2}{3}\,(r/r_x)^2{\cal F}(r/r_x)\,{\rm e}^{-r_i/r_x}
\end{eqnarray}
if $r_x > r$ (the external shells). Given these expressions, we can (and do) integrate to get the total force at a radius $r$.  

To get a qualitative feel for the effect of $\alpha\ne 0$, it is instructive to study the net force associated with a tophat perturbation, for which $\delta_i$ is the same for all shells within the perturbation.  (Note that all the analysis in the main text, and the figures in this Appendix, were made by solving for the exact evolution, i.e., not assuming that the initial tophat remains a tophat.)  In this case the modification to the Newtonian force $-GM/r^2$ is given by 
\begin{eqnarray}
 F_\alpha(r) &=& \alpha \frac{G\delta M}{r^2}\,
              \Bigl[{\rm e}^{-r/r_s}\,(1+r/r_s)\, {\cal F}(R/r_s) \\
    && \qquad\qquad\quad - {\rm e}^{-r/r_c}\,(1+r/r_c)\, {\cal F}(R/r_c)\Bigr]
         \nonumber\\
  &\approx& -\alpha\, \frac{G\delta M}{r^2}\,
              \Bigl[1 - {\rm e}^{-r/r_s}\,(1+r/r_s)\, {\cal F}(R/r_s)\Bigr].
         \nonumber
\end{eqnarray}
In the present context, we are also interested in the profile within the perturbation.  A similar analysis of $\Phi(r)$ when $r<R$ shows that the modification to the force is 
\begin{eqnarray}
 F_\alpha(r) &=& \alpha \frac{G\,\delta M(<r)}{r^2} 
   \Bigl[{\rm e}^{-R/r_s}(1+R/r_s)\,{\cal F}(r/r_s)\nonumber\\
        && \qquad\qquad - {\rm e}^{-R/r_c}(1+R/r_c)\,{\cal F}(r/r_c)\Bigr]\\
   &\approx& -\alpha \frac{G\,\delta M(<r)}{r^2} 
         \Bigl[1-{\rm e}^{-R/r_s}(1+R/r_s)\,{\cal F}(r/r_s)\Bigr].\nonumber
\end{eqnarray}
The term in square brackets is always positive, so the correction is proportional to the product of $\alpha$ and $\delta M$.  Hence, when both are positive, then the resulting profile becomes steeper than when $\alpha=0$:  one might generically expect halos to be more concentrated.  However, for overdense perturbations when $\alpha<0$, the net force towards the center is smaller than when $\alpha=0$, so we expect the profile to evolve away from a tophat in the sense of becoming more concentrated, not at the center, but at the outer boundary.  

We have looked for this effect in the simulations we study in the main text.  Unfortunately, because the particle mass in these simulations is rather large (the runs were designed to study larger scale structures), we can only estimate halo density profiles for the rarest most massive objects.  In the $\alpha = (-1,0,0.5)$ runs, the most massive object had $\log_{10} (M/h^{-1}M_\odot) = (14.85, 15.15, 15.58)$ and virial radii, $R_{\rm vir}/h^{-1}{\rm Mpc} = (1.31, 1.66, 2.36)$.  Figure~\ref{simprofs} shows the profiles of these three objects.  In each case, we show the expected enclosed density profile for an object of this mass if $\alpha = 0$.  In all cases, the measured profiles lie below this line on scales smaller than $R_{\rm vir}$, suggesting that our simulations are not able to properly resolve small scale structures.  However, there is a hint that the $\alpha = -1$ object has slightly more than the expected amount of mass just beyond the virial radius, whereas the opposite is true for the $\alpha = 0.5$ object.  It will be interesting if these trends persist in simulations with better mass and force resolution, because it is conceivable that effects of this magnitude will soon be measured by weak lensing surveys.  

Finally, we note that there is another sense in which the modified gravity model is more complicated.  As the object pulls itself together from the expansion, it will eventually reach turnaround.  In standard gravity, all its shells reach turnaround at the same time, so the total energy at turnaround is potential, and energy conservation means that this equals its initial energy.  When $\alpha\ne 0$, however, the different shells turnaround at different times, so the total energy when the outermost shell turns around is no longer all potential. 

\begin{figure}[t]
\includegraphics[width=0.95\hsize]{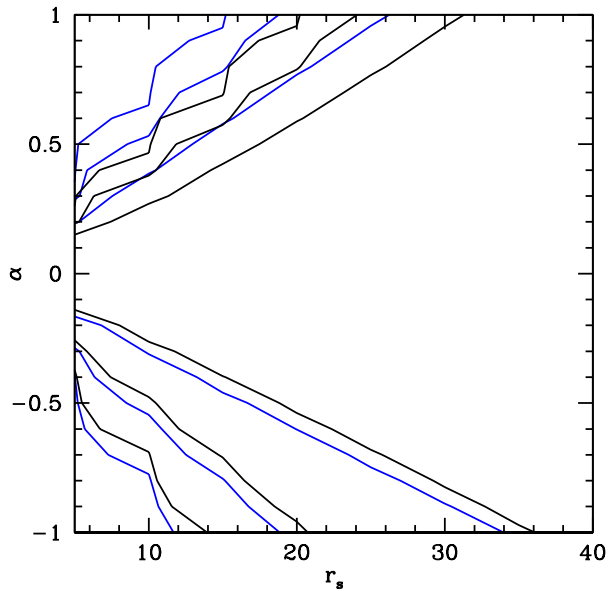}
\caption{One, two, and three $\sigma$ confidence intervals in the $(\alpha,r_s)$ plane when the initial power-spectra are the same (blue) and when the $z=0$ linear theory power spectra are the same (black).  Constraints come from requiring the models to agree with the cluster abundances measured by \cite{vikhlinin}, and assuming that the scaling relation between cluster observable and mass is the same as when $\alpha =0$. }
\label{fig:contours}
\end{figure}

\section{Constraining $\alpha$ and $r_s$}\label{alphars}
 The main text studied a particular modification to large-scale gravity, which has two free parameters:  a scale $r_s$, and an amplitude $\alpha$.  Previous work has sought to constrain the values of these parameters by using the shape of the galaxy power spectrum \cite{shirata1,shirata2}.  Although, on large scales, the amplitude of the power-spectrum depends on $\alpha$, potentially providing a powerful constraint, this is offset by the fact that galaxies are biased tracers of the dark matter field, and this bias is unknown.  Even allowing for the simplest, linear bias relation removes most of the sensitivity of this method to the information contained in the amplitude of the late-time $P(k)$.  Our succesful model for halo abundances, however, is sensitive to precisely this difference, so, we can use current high mass cluster counts to constrain the allowed range of $(\alpha,r_s)$.  


Figure~\ref{fig:contours} shows the result of comparing our predicted mass functions with the measured counts of \cite{vikhlinin} when we normalize to the same (CMB) initial conditions for all $\alpha$ (blue contours), and when normalized to have the same linear theory $P(k)$ at $z=0$ (black).  The inner most contours represent one, two, and three $\sigma$ confidence levels.  The constraints for the CMB normalized curves are slightly weaker than the others.  This can be traced to Figure~6 in \cite{mss09}, which shows that the solid line is closer to unity than the dotted line in the mass range of clusters, $10^{14}-10^{15}\,M_\odot/h$.  This is because the mass function is determined by the shape of the barrier and the shape of the \emph{initial} power spectrum, so that in the case of CMB normalization only one of these components changes (the barrier), whereas for cluster normalization both change, leading to a larger change in the mass function -- and hence to stronger constraints.  Note in particular, that the constraints that we derive here are roughly comparable to those derived by \cite{shirata1,shirata2}.  However, if halo density profiles are indeed sensitive to $\alpha$, as our analysis suggests, then this will affect the scaling relations that are usually assumed to convert X-ray observables into masses.  This, in turn, will alter the derived constraints.  We look forward to the day when better simulations are available, so the question of how X-ray cluster scaling relations depend on $\alpha$ and $r_s$ has been settled.

\label{lastpage}

\end{document}